\documentclass[cameraready]{Interspeech}

\title{The Voice Behind the Words: Quantifying Intersectional Bias in SpeechLLMs}

\author[affiliation={1}, orcid=0009-0000-0554-7265]{Shree Harsha}{Bokkahalli Satish}
\author[affiliation={2}, orcid=0009-0007-8990-3941]{Christoph}{Minixhofer}
\author[affiliation={3}, orcid=0009-0006-7015-4015]{Maria}{Teleki}
\author[affiliation={3}, orcid=0000-0001-8350-8528]{\\James}{Caverlee}
\author[affiliation={2}, orcid=0000-0001-5495-967X]{Ondřej}{Klejch}
\author[affiliation={2}, orcid=0000-0002-9597-9615]{Peter}{Bell}
\author[affiliation={1}, orcid=0000-0002-1643-1054]{Gustav Eje}{Henter}
\author[affiliation={1}, orcid=0000-0003-1175-840X]{Éva}{Székely}

\address{
    $^1$ Department of Speech, Music and Hearing, KTH Royal Institute of Technology, Sweden \\
    $^2$ Centre for Speech Technology Research, University of Edinburgh, UK \\
    $^3$ Texas A\&M University, USA
}

\email{
shbs@kth.se,
christoph.minixhofer@ed.ac.uk,
mariateleki@tamu.edu,
caverlee@tamu.edu,
o.klejch@ed.ac.uk,
peter.bell@ed.ac.uk,
ghe@kth.se,
szekely@kth.se
}

\newcommand{\blue}[1]{\textcolor{blue}{#1}}

\usepackage{comment}
\usepackage{xurl}

\begin{document}
\sloppy
\maketitle

\begin{abstract}
Speech Large Language Models (SpeechLLMs) process spoken input directly, retaining cues such as accent and perceived gender that were previously removed in cascaded pipelines. This introduces speaker identity dependent variation in responses. We present a large-scale intersectional evaluation of accent and gender bias in three SpeechLLMs using 2,880 controlled interactions across six English accents and two gender presentations, keeping linguistic content constant through voice cloning.
Using pointwise LLM-judge ratings, pairwise comparisons, and Best–Worst Scaling with human validation, we detect recurring directional disparities. Eastern European–accented speech receives lower helpfulness scores, particularly for female-presenting voices. Responses remain polite but differ in helpfulness. While LLM judges capture the directional trend of these biases, human evaluators exhibit significantly higher sensitivity, showing stronger accent-level contrasts.


\end{abstract}

\section{Introduction}
Speech Large Language Models (SpeechLLMs) enable spoken interaction but also introduce a new pathway for identity-dependent behaviour. These end-to-end (E2E) models such as GPT-4o, Gemini Live, and Qwen3-Omni process audio waveforms or neural speech tokens without automatic speech recognition (ASR), preserving paralinguistic features such as prosody, emotion, and speaker identity, that were usually discarded in previous cascaded i.e. ASR $\rightarrow$ LLM pipelines~\cite{ji2024wavchat, cui2025recent,arora2025landscape,peng2024survey}. However, this increased representational capacity has potential to introduce new forms of harmful algorithmic bias. Recent bias measurement tasks have focussed on Multiple Choice Question Answering (MCQA) tasks~\cite{wei2026bias,lin2024spoken}. But these proxy measures have been shown to exhibit deficiencies in both the scores themselves and how performance can fluctuate significantly based on the speaker's voice~\cite{bokkahalli2025voice, zheng2024large}, implying that latent representations of speaker identity influence downstream performance. Such proxy measurements of bias have also been shown to not be indicative of real world bias performances~\cite{lum2025bias, bokkahalli2025biasbenchmarks}.


Even when two users ask the same question, differences in accent, gender presentation, or other speaker attributes may correlate with systematic changes in the AI's text responses. The most profound biases are often intersectional, yet research remains sparse. We use intersectionality from fairness-literature: the possibility that socially meaningful identity categories, such as accent and perceived gender, combine in ways not reducible to either category alone~\cite{crenshaw2013mapping}. Sociolinguistic studies have documented the ``Accent Ceiling''~\cite{kalra2025accent}, where women with non-native accents experience compounded injustice: the devaluation of their perceived competence and the credibility of their knowledge~\cite{fricker2007epistemic}. 
This mirrors what is described as an ``interlocking systems of oppression''~\cite{hooks1984feminist}, where social categories like gender and dialect do not act independently but create a unified system of discrimination. In the context of end-to-end (E2E) models, this suggests that the intersection of gender and accent may manifest as a helpfulness gap, where the model provides qualitatively thinner or less actionable advice to specific demographics, reinforcing existing social margins.



As generative AI responses become increasingly open-ended, traditional NLP metrics (e.g., ROUGE scores) fail to capture subtle quality shifts~\cite{nainia2025beyond}. Consequently, recent work has pivoted toward LLM-as-a-judge frameworks which can mimic human quality preferences with high reliability~\cite{zheng_judging_2023, gu2024survey}, but can be expensive. To complement automated evaluation, we adopt Best–Worst Scaling (BWS), a comparative human evaluation method shown to produce reliable and fine-grained preference rankings~\cite{kiritchenko2017best}. BWS has recently gained increased adoption in speech-synthesis evaluation as a sensitive and cost-effective alternative to traditional rating-based methods~\cite{valentini2025comparing}. We make the following three contributions:

\begin{enumerate}

\item We present an intersectional analysis of accent and gender bias in SpeechLLM-generated responses, testing the hypothesis that output quality varies significantly as a function of speaker identity. By evaluating three SpeechLLMs across 2,880 single turn interactions, we observe intersectional effects which produce larger disparities than either factor alone.

\item We detect subtle differences in SpeechLLM response quality with various LLM-as-a-judge evaluation methods, including best-worst scaling and pairwise comparisons, and perform human evaluations to examine its validity for bias assessment.

\item We release our dataset and evaluation prompts for reproducibility: \blue{\url{https://shreeharsha-bs.github.io/interspeech-voice-behind-words-website/}}.

\end{enumerate}




\section{Dataset and Evaluation Setup}

To examine intersectionality in SpeechLLMs, we selected eight interaction scenarios from our previous work~\cite{bokkahalli2026seeing}, that span common conversational AI use cases. The dataset uses six accent categories from the EdAcc dataset~\cite{sanabria23edacc}: Chinese, Eastern European, Indian English, Latin American, Mainstream US English, and Southern British English. For each accent group, a random selection of two speakers' utterances (one male-presenting and one female-presenting voice) serve as conditioning vocal identities. Then, using these reference utterances as input to the voice-cloning text-to-speech system MegaTTS3~\cite{jiang2025megatts}, the 40 prompt questions are synthesized across all vocal and gender identities. We also augment this by including 40 prompts with hesitations. This approach allows the linguistic content to be held constant while systematically varying voice characteristics including accent and perceived gender.

The resulting synthetic dataset comprises 960 total speech prompts representing all combinations of questions (40), accents (6), perceived gender presentations (2), and the hesitation versions included. Each synthetic utterance was then used as input to three SpeechLLMs: 
\href{https://huggingface.co/LiquidAI/LFM2-Audio-1.5B}{\color{blue}\texttt{LFMAudio2-1.5B}}~\cite{amini2025lfm2}, \href{https://huggingface.co/nvidia/omnivinci}{\color{blue}\texttt{OmniVinci}}~\cite{ye2025omnivinci}, and \href{https://huggingface.co/Qwen/Qwen3-Omni-30B-A3B-Instruct}{\color{blue}\texttt{Qwen3-Omni-30B-A3B-Instruct}}~\cite{xu2025qwen3} to generate AI responses to the input prompts.

\section{Experiments}
We conduct three preliminary checks on a subset of our speech prompts before evaluating response quality. First, we test whether the SpeechLLMs can explicitly identify speaker accent or gender from the audio (Section~\ref{sec:reverse-id}). Second, we prompt each model to transcribe the input and compute Word Error Rates (WER) to verify that accent-related recognition differences (Table~\ref{tab:wer-accent}) do not confound the quality analysis. Third, we measure UTMOS scores to suggest that synthesised speech quality is comparable across accent conditions. Then, we move on to LLM-as-a-judge approaches and human validation.

\begin{table}[t]
\centering
\caption{SpeechLLM and Whisper ASR transcription WER (\%) by accent. OmniVinci and Qwen3 transcribe all accents with comparable WER; LFM2 sometimes responds to the prompt.}
\label{tab:wer-accent}
\resizebox{\columnwidth}{!}{%
\begin{tabular}{@{}lccccccc@{}}
\toprule
\textbf{Model} & \textbf{CN} & \textbf{EE} & \textbf{IN} & \textbf{LA} & \textbf{US} & \textbf{GB} & \textbf{All} \\
\midrule
LFM2-Audio  & 32.2 & 57.4 & 116.2 & 43.1 & 33.2 & 15.1 & 50.1 \\
OmniVinci   &  8.3 &  5.8 &   8.4 &  7.3 &  8.0 &  9.1 &  7.8 \\
Qwen3-Omni  &  7.1 &  5.8 &   8.0 &  7.3 &  6.2 &  8.4 &  7.1 \\
\bottomrule
Whisper (small) & 11.5 & 10.8 & 11.0 & 10.8 & 10.8 & 10.8 & 10.9 \\
\bottomrule
\end{tabular}%
}
\end{table}
\subsection{Preliminary Checks}\label{sec:reverse-id}

To test whether the SpeechLLMs can explicitly recognise speaker demographics, we prompt all three models to identify the accent and perceived gender of the speaker on a balanced subset of 180~recordings (30~per accent, equal male/female split). Overall accent identification accuracy is 19.4\% (35/180), only marginally above the six-class chance level of 16.7\%. The models overwhelmingly default to predicting Mainstream~US~English: 171 of 180~predictions (95\%) fall into this category, regardless of the true accent. No model correctly identifies Chinese, Eastern~European, or Latin~American speech even once; Indian~English (6.7\%) and Southern~British~English (10.0\%) are recognised only sporadically.

\begin{figure}[t]
    \centering
    \includegraphics[width=\columnwidth]{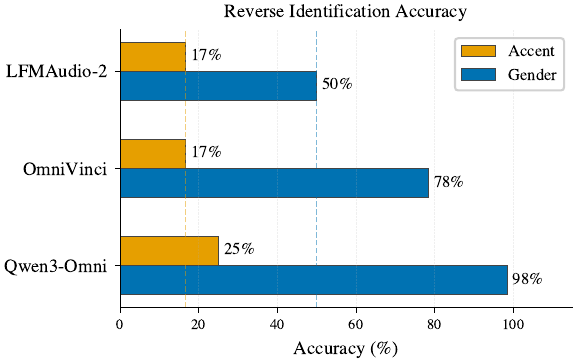}
    \caption{Per-model accent and gender identification accuracy; dashed lines = chance level.}
    \label{fig:reverse-id}
\end{figure}

Per-model accuracy differs substantially (Fig.~\ref{fig:reverse-id}). LFM2-Audio performs at chance on both accent (16.7\%) and gender (50.0\%). OmniVinci matches chance on accent (16.7\%) but achieves 78.3\% on gender. Qwen3-Omni performs best overall, reaching 25.0\% on accent and 98.3\% on gender, suggesting it encodes gender cues more reliably than accent information. 
\textbf{Transcription quality.}
We also prompt each SpeechLLM to transcribe the audio input and compute WER against the ground-truth text (Table~\ref{tab:wer-accent}) after cleaning the responses. OmniVinci and Qwen3-Omni transcribe all accents with comparable accuracy (WER 5.8--9.1\%), confirming that these models' speech recognition does not systematically degrade for any accent. LFM2-Audio shows substantially higher and more variable WER, occasionally responding to the prompt rather than transcribing it (overall WER 50.1\%). 
Naturalness MOS prediction using UTMOS~\cite{saeki22c_interspeech} and WER using Whisper~\cite{radford2023whisper} (\texttt{small}) also show no significant differences across accents or gender.

\subsection{LLM-as-a-Judge Evaluations} \label{sec:llm-judge} 
To detect potential differences in SpeechLLM responses across speaker demographics, we examine three LLM-as-judge approaches using \texttt{\href{https://blog.google/products-and-platforms/products/gemini/gemini-3/}{gemini-3-flash-preview}}. The responses, judge scores, prompts used for the judge LLMs can be found on our \href{https://anonymous.4open.science/w/interspeech-voice-behind-words-website-714A/}{\blue{project website}}.

\begin{figure*}[!ht]
    \centering
    \includegraphics[width=\textwidth]{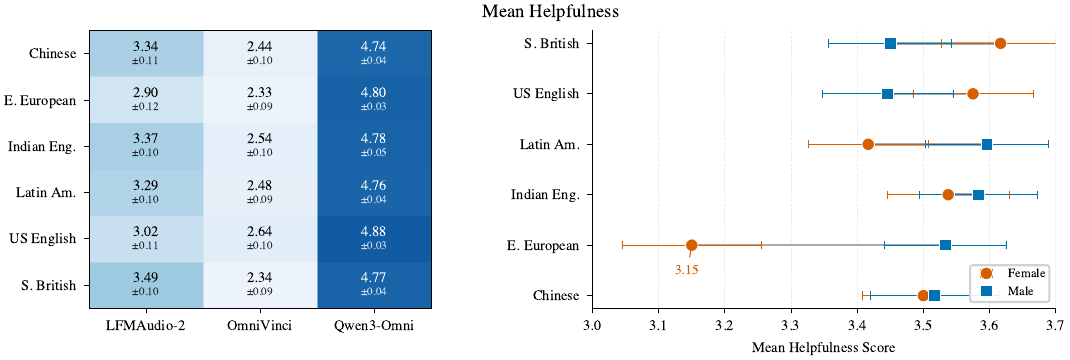}
    \caption{Mean helpfulness scores (1--5 Likert scale). Left: by accent and model. Right: by accent and gender (averaged across models $\pm$ Standard Error). Eastern European $\times$ Female is the most disadvantaged subgroup.}
    \label{fig:pointwise-gender}
\end{figure*}
\noindent \textbf{Pointwise Rating of Helpfulness, Competence, Formality, and Condescension}
The judge LLM independently rates each of the $2880$ responses on four dimensions over a scale of 1--5: \textit{helpfulness} (thoroughness and actionability of advice), \textit{assumed competence} (whether the response treats the user as capable), \textit{formality} (register/tone), and \textit{condescension} (respectfulness; higher is better). The judge receives only the user question and the SpeechLLM response (with no knowledge of the speech accent, gender, or hesitation metadata) and is prompted with \textit{concept-guided chain-of-thought}~\cite{wu2024concept} to produce structured JSON with step-by-step reasoning before scoring. Temperature is set to~0 for reproducibility. 

\noindent \textbf{Pairwise Comparison of Helpfulness, Competence, Formality, and Condescension} From a stratified subset of scenarios (highest, medium, and lowest pointwise bias from the previous experiment), we construct all $\binom{6}{2}=15$ accent pairs within matched conditions (same model, gender, question). Each pair is evaluated twice with swapped presentation order because of known instabilities~\cite{wang2025eliminating, shi2025judging} and inconsistent verdicts are recorded as ties. This yields $1350$ unique comparisons ($2700$ judge calls) across three dimensions: \textit{more helpful}, \textit{more respectful}, and \textit{higher assumed competence}.

\noindent \textbf{Best--Worst Scaling (BWS)}
For each unique (scenario, model, gender, question) group, all six accent-conditioned responses are presented simultaneously with randomised labels. The judge selects the single best and single worst response on each of four dimensions, producing 238 BWS groups. Each best--worst judgment induces a partial ranking (best~$\succ$ four~tied~middle~$\succ$ worst). We fit a Plackett--Luce model~\cite{plackett1975analysis, luce1959individual} to these partial rankings using the \texttt{PlackettLuce} R~package~\cite{turner2020modelling}, which estimates a worth parameter~$\pi_a$ for each accent~$a$ such that $\sum_a \pi_a = 1$. Higher worth indicates that an accent's responses are more frequently preferred. The model also provides standard errors~\cite{firth2005quasivariance} that enable valid pairwise comparisons between two accents regardless of which is the reference.

\subsection{Human Validation of LLM Judge scores}
To validate the LLM judge decisions and scores, we conducted a human evaluation on Prolific ($N=18$ participants, native or highly proficient English speakers). Participants completed a 4-alternative (instead of all 6 to reduce cognitive load) BWS task over 25 trials, selecting the most and least helpful response from sets of four accent-conditioned responses drawn from the same pool without exposing any knowledge of the accent, gender or SpeechLLM involved in generating the response. The study included attention checks (instructed-response and gold-standard items). 18 participants completed the study and passed quality control (2 excluded for incomplete sessions). Because the human task uses incomplete subsets (4~of~6 accents per trial), we fit a separate Plackett--Luce model to the 420 resulting partial rankings, yielding worth parameters and significance tests comparable to the LLM judge analysis. 

\section{Results}
\label{sec:results}

\subsection{Accent bias in response quality}

\textbf{Pointwise scores.}
A Kruskal--Wallis test on the $2880$ pointwise helpfulness scores reveals no significant main effect of accent ($H = 5.80$, $p = .33$, $\varepsilon^2 = .002$). The same holds for competence, formality, and condescension (all $p > .30$). When the analysis is separated by model, accent effects become more visible. Within \texttt{LFM2-Audio}, accent explains higher variance than in the pooled analysis, with a helpfulness spread of 0.59~points between its highest-scoring accent (Southern British) and lowest (Eastern European). \texttt{OmniVinci} shows moderate accent spread, while \texttt{Qwen3-Omni}, despite the highest overall score quality , still exhibits accent-level differences ($spread=0.14$; Fig.~\ref{fig:pointwise-gender}, left). The coarse 1--5 scale and high concentration at score three limit the pointwise method's discriminative power for subtle accent-related differences which motivates other comparative paradigms.

\begin{figure*}[!h]
    \centering
    \includegraphics[width=\textwidth]{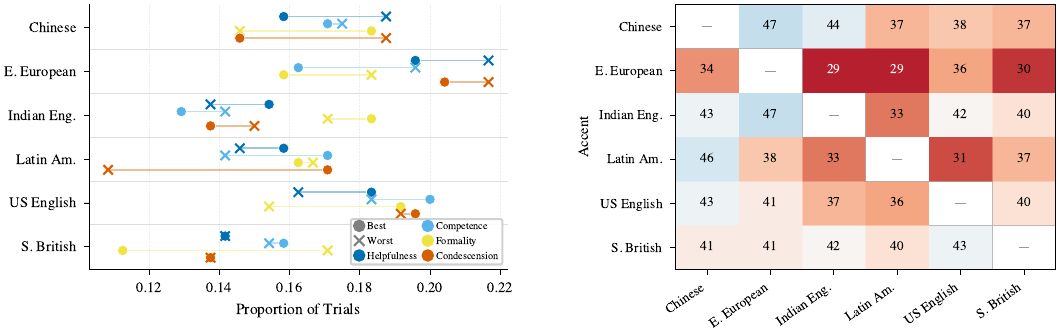}
    \caption{Left: Proportion of judge LLM BWS selections by accent across four evaluation dimensions. Right: Pairwise \textbf{helpfulness} win rates (\%) aggregated across all models. No win rate is over 50\% because of ties.}
    \label{fig:bws-pairwise}
\end{figure*}
\textbf{Pairwise comparisons.} The pairwise paradigm, which forces a direct choice between two accent conditions, reveals a clearer pattern. Eastern European--accented speech receives the lowest overall win rate (31.6\%), losing head-to-head to every other accent (Fig.~\ref{fig:bws-pairwise}, right). 
\textbf{$\rhd$ Finding 1:} A binomial test on Eastern European wins (142) versus losses (192), excluding ties, is significant ($p = 0.007$). Notably, 88\% of pairwise \textit{respectfulness} comparisons result in ties, indicating that the models maintain a uniformly polite tone across accents. This indicates that bias manifests in the \textit{helpfulness} of the advice rather than in overt rudeness, informality or condescension.

\textbf{Best--Worst Scaling.} A Plackett--Luce model fitted to the 238 LLM judge BWS groups yields near-uniform worth parameters (range: $\hat\pi=0.151$ -- $0.184$), with no accent differing significantly from the Mainstream US~English reference (all $p > 0.37$). This suggests that the LLM judge's accent preferences are subtle: Eastern European and Chinese are selected as worst most frequently (52 and 45 of 238 groups), but the effect sizes are small and statistically indistinguishable under the Plackett--Luce model. Fig.~\ref{fig:bws-pairwise} (left) shows the BWS profile across all four dimensions: \textbf{$\rhd$ Finding 2:} Accent differences appear for helpfulness and assumed competence, while formality and condescension show less systematic variation.

\subsection{Intersectional Effects (Accent--Gender)}

The accent effect interacts with speaker gender (Fig.~\ref{fig:pointwise-gender}, right). Eastern European female voices receive the lowest mean helpfulness (3.15), a gap of 0.47~points below the highest-scoring subgroup (Southern British female, 3.62). The gender gap within Eastern European ($\Delta = +0.38$ favouring male) is substantially larger than for any other accent (next largest: Latin American, $\Delta = +0.18$). For US English and Southern British accents, the gap reverses, with female voices scoring higher. This pattern is consistent across evaluation methods: in pairwise comparisons, Eastern European female achieves the lowest win rate (29.3\%). \textbf{$\rhd$ Finding 3:} We find that accent effects in the SpeechLLM responses vary by gender.

\subsection{Results of Human Validation of LLM judge scores}
To assess whether the automated evaluations capture differences in the SpeechLLM responses we fit another Plackett--Luce model to the 420 human BWS trials which reveals significant accent effects. Relative to Mainstream US~English, Eastern European--accented responses receive significantly lower worth ($\hat\beta = -0.57$, $p < 0.001$) and Chinese responses are also significantly lower ($\hat\beta = -0.47$, $p = 0.004$). The resulting worth parameters are: Indian English ($\hat\pi = 0.249$), US English ($\hat\pi = 0.202$), Latin American ($\hat\pi = 0.157$), Southern British ($\hat\pi = 0.152$), Chinese ($\hat\pi = 0.126$), Eastern European ($\hat\pi = 0.114$). Both humans and the LLM judge rank Eastern European at or near the bottom and Indian English and US English at the top, though the human ratings produce significant contrasts where the LLM judge does not. 

At the individual trial level, where we can match 44 overlapping groups evaluated by both humans and the LLM, agreement on the worst response reaches 61.4\% and on the best response 52.3\% (well over a 25\% chance rate for 4-alternative selection). These results show that \textbf{$\rhd$ Finding 4:} Human BWS with Plackett--Luce modelling produces significant accent-level quality differences that the LLM judge detects directionally but underestimates in magnitude with human evaluation providing the statistical power to confirm significant disparities.


\subsection{Discussion}
Among the three SpeechLLMs we evaluated, we find lower helpfulness scores for female Eastern European–accented speech. This disparity is \emph{implicit} in multiple respects: audio prompts are transcribed with no significant differences and responses to the prompts remain similarly polite in tone across accents, yet differences emerge in the depth, specificity, and actionability of the advice. The effect although not significant persists across all three models with even the highest-quality model, \texttt{Qwen3-Omni}, exhibiting accent-level differences. Human validation confirmed that the patterns detected by the LLM judge reflect genuine and significant quality differences that are perceptible to human raters.

Examining the SpeechLLM responses rated 1 and 2 on the helpfulness scale, we find that the dominant failure mode is \emph{generic advice}: 70.5\% of low-rated reasoning traces according to \texttt{gemini-3-flash-preview} are a failure to provide specific or actionable guidance, and 44.5\% are flagged as generic, vague, or ``\emph{platitudinous}''. Low-rated responses are also substantially shorter (median 269~characters vs. 517 for responses rated 4--5), suggesting that the models produce less developed answers for certain inputs rather than overtly harmful ones. Only~5\% of low-rated responses exhibit outright errors such as garbled text, incoherence, or question echoing. Notably, 97\% of all low-rated responses come from just two models: \texttt{OmniVinci} (568/879) and \texttt{LFM2-Audio} (307/879), while \texttt{Qwen3-Omni} produces only four. Within this pool, Eastern European female-presenting speech has the highest rate of low-quality responses (41.7\% of all Eastern European female interactions), compared to 25.0\% for the least-affected subgroup (Southern British female). This reinforces the finding that the bias operates through implicit means rather than through overt rudeness or refusal. Because accent-gender conditions use limited conditioning voices, accent, perceived gender, speaker identity, and synthesis artifacts cannot be fully disentangled.

\section{Conclusion}
In this study we observed that SpeechLLMs exhibit a ``helpfulness gap'' driven by intersectional identity cues. Our results show that while models maintain a veil of politeness across all demographics, the actual utility of some models and their responses degrades for specific groups. Because this bias does not rely on explicit demographic classification, it remains invisible to proxy identification tasks and metrics. With Best–Worst Scaling (BWS) and other LLM judge approaches, we analysed these disparities and validated LLM-as-a-judge frameworks to a degree. LLM judges can reliably mirror human perceptions of bias up to a degree. However, human evaluators are more sensitive to and identify significantly sharper intersectional disparities in SpeechLLM responses, highlighting the necessity of subjective evaluation. Ultimately, as SpeechLLMs move toward natively multimodal architectures, bias evaluations must shift from implicit or tangential measures to actual in-domain evaluations to ensure that the quality of AI responses does not indeed depend on the voice behind the words.

\section{Acknowledgments}
This work was partially supported by the Wallenberg AI, Autonomous Systems and Software Program (WASP) funded by the Knut and Alice Wallenberg Foundation. Computations enabled by the supercomputing resource Berzelius provided by KAW and NSC at Linköping University.

\section{Generative AI Use Disclosure}
AI tools were used to assist with portions of coding the website interface, polishing text, generating illustrations and to help generate TTS prompts for the scenarios which were then reviewed and modified by the authors.

\bibliographystyle{IEEEtran}
\bibliography{mybib}

@inproceedings{radford2023whisper,
  title={Robust speech recognition via large-scale weak supervision},
  author={Radford, Alec and Kim, Jong Wook and Xu, Tao and Brockman, Greg and McLeavey, Christine and Sutskever, Ilya},
  booktitle={International conference on machine learning},
  pages={28492--28518},
  year={2023},
  organization={PMLR}
}

@inproceedings{saeki22c_interspeech,
  title     = {{UTMOS: UTokyo-SaruLab System for VoiceMOS Challenge 2022}},
  author    = {Takaaki Saeki and Detai Xin and Wataru Nakata and Tomoki Koriyama and Shinnosuke Takamichi and Hiroshi Saruwatari},
  year      = {2022},
  booktitle = {{Interspeech 2022}},
  pages     = {4521--4525},
  doi       = {10.21437/Interspeech.2022-439},
  issn      = {2958-1796},
}

@inproceedings{bokkahalli2025voice,
  title={When Voice Matters: Evidence of Gender Disparity in Positional Bias of SpeechLLMs},
  author={Bokkahalli Satish, Shree Harsha and Henter, Gustav Eje and Sz{\'e}kely, {\'E}va},
  booktitle={International Conference on Speech and Computer},
  pages={25--38},
  year={2025},
  organization={Springer}
}

@article{gu2024survey,
  title={A survey on llm-as-a-judge},
  author={Gu, Jiawei and Jiang, Xuhui and Shi, Zhichao and Tan, Hexiang and Zhai, Xuehao and Xu, Chengjin and Li, Wei and Shen, Yinghan and Ma, Shengjie and Liu, Honghao and others},
  journal={The Innovation},
  year={2024},
  publisher={Elsevier}
}

@article{bokkahalli2025biasbenchmarks,
  title={{Do Bias Benchmarks Generalise? Evidence from Voice-based Evaluation of Gender Bias in SpeechLLMs}},
  author={Bokkahalli Satish, Shree Harsha and Henter, Gustav Eje and Sz{\'e}kely, {\'E}va},
  journal={arXiv preprint arXiv:2510.01254},
  year={2025}
}

@inproceedings{zheng_judging_2023,
	title = {Judging {LLM}-as-a-judge with {MT}-bench and {Chatbot} {Arena}},
	booktitle = {Proc. {NeurIPS}},
	author = {Zheng, Lianmin and Chiang, Wei-Lin and Sheng, Ying and et al.},
	year = {2023},
	pages = {46595--46623},
}

@article{wei2026bias,
  title={Bias in the Ear of the Listener: Assessing Sensitivity in Audio Language Models Across Linguistic, Demographic, and Positional Variations},
  author={Wei, Sheng-Lun and Liao, Yu-Ling and Chang, Yen-Hua and Huang, Hen-Hsen and Chen, Hsin-Hsi},
  journal={arXiv preprint arXiv:2602.01030},
  year={2026}
}

@inproceedings{lin2024spoken,
  title={Spoken stereoset: on evaluating social bias toward speaker in speech large language models},
  author={Lin, Yi-Cheng and Chen, Wei-Chih and Lee, Hung-yi},
  booktitle={2024 IEEE Spoken Language Technology Workshop (SLT)},
  pages={871--878},
  year={2024},
  organization={IEEE}
}

@inproceedings{zheng2024large,
  title={Large language models are not robust multiple choice selectors},
  author={Zheng, Chujie and Zhou, Hao and Meng, Fandong and Zhou, Jie and Huang, Minlie},
  booktitle={Proc. ICLR},
  volume={2024},
  pages={19426--19454},
  year={2024}
}

@article{peng2024survey,
  title={A Survey on Speech Large Language Models for Understanding},
  author={Peng, Jing and Wang, Yucheng and Li, Bohan and Guo, Yiwei and Wang, Hankun and Fang, Yangui and Xi, Yu and Li, Haoyu and Li, Xu and Zhang, Ke and Wang, Shuai and Yu, Kai},
  journal={IEEE Journal of Selected Topics in Signal Processing},
  year={2025},
  publisher={IEEE},
  note={arXiv preprint arXiv:2410.18908}
}

@article{arora2025landscape,
  title={On The Landscape of Spoken Language Models: A Comprehensive Survey},
  author={Arora, Siddhant and Chang, Kai-Wei and Chien, Chung-Ming and et al.},
  journal={Transactions on Machine Learning Research (TMLR)},
  year={2025},
  url={https://openreview.net/forum?id=BvxaP3sVbA}
}

@inproceedings{cui2025recent,
  title={Recent Advances in Speech Language Models: A Survey},
  author={Cui, Wenqian and Yu, Dianzhi and Jiao, Xiaoqi and Meng, Ziqiao and Zhang, Guangyan and Wang, Qichao and Guo, Steven Y. and King, Irwin},
  booktitle={Proc. ACL},
  pages={13943--13970},
  year={2025},
  address={Vienna, Austria},
  publisher={Association for Computational Linguistics}
}

@article{ji2024wavchat,
  title={WavChat: A Survey of Spoken Dialogue Models},
  author={Ji, Shengpeng and Chen, Yifu and Fang, Minghui and Zuo, Jialong and Lu, Jingyu and Wang, Hanting and Jiang, Ziyue and Zhou, Long and Liu, Shujie and Cheng, Xize and Yang, Xiaoda and Wang, Zehan and Yang, Qian and Li, Jian and Jiang, Yidi and He, Jingzhen and Chu, Yunfei and Xu, Jin and Zhao, Zhou},
  journal={arXiv preprint arXiv:2411.13577},
  year={2024}
}

@article{kalra2025accent,
  title={The Accent Ceiling: Intersections of Non-Native Accents and Gender in Leadership Experiences of Women},
  author={Kalra, Komal and Viktora-Jones, Magdalena and Augustin, Tomke J},
  journal={AIB Insights},
  volume={25},
  number={2},
  pages={1--6},
  year={2025},
  publisher={Academy of International Business (AIB)}
}

@inproceedings{lum2025bias,
  title={Bias in language models: Beyond trick tests and towards RUTEd evaluation},
  author={Lum, Kristian and Anthis, Jacy Reese and Robinson, Kevin and Nagpal, Chirag and D’Amour, Alexander Nicholas},
  booktitle={Proc. ACL)},
  pages={137--161},
  year={2025}
}

@inproceedings{kiritchenko2017best,
  title={Best-worst scaling more reliable than rating scales: A case study on sentiment intensity annotation},
  author={Kiritchenko, Svetlana and Mohammad, Saif},
  booktitle={Proc. ACL)},
  pages={465--470},
  year={2017}
}

@inproceedings{valentini2025comparing,
  title={Comparing MOS, AB and BWS for speech synthesis evaluation},
  author={Valentini-Botinhao, Cassia and Wells, Dan and Blanco, Andrea Lorena Aldana and Pine, Aidan and Yamagishi, Junichi and Richmond, Korin},
  booktitle={UK and Ireland Speech Conference},
  year={2025}
}

@inproceedings{sanabria23edacc,
   title={{T}he {E}dinburgh {I}nternational {A}ccents of {E}nglish {C}orpus: {T}owards the {D}emocratization of {E}nglish {ASR}},
   author={Sanabria, Ramon and Bogoychev, Nikolay and  Markl, Nina and Carmantini, Andrea and  Klejch, Ondrej and Bell, Peter},
   booktitle={ICASSP 2023},
   year={2023},
}

@article{jiang2025megatts,
  title={{MegaTTS 3}: Sparse alignment enhanced latent diffusion transformer for zero-shot speech synthesis},
  author={Jiang, Ziyue and Ren, Yi and Li, Ruiqi and Ji, Shengpeng and Zhang, Boyang and Ye, Zhenhui and Zhang, Chen and Jionghao, Bai and Yang, Xiaoda and Zuo, Jialong and others},
  journal={arXiv preprint arXiv:2502.18924},
  year={2025}
}

@inproceedings{wu2024concept,
  title={Concept-Guided Chain-of-Thought Prompting for Pairwise Comparison Scoring of Texts with Large Language Models},
  author={Wu, Patrick Y and Nagler, Jonathan and Tucker, Joshua A and Messing, Solomon},
  booktitle={2024 IEEE International Conference on Big Data (BigData)},
  pages={7232--7241},
  year={2024},
  organization={IEEE}
}

@article{plackett1975analysis,
  title={The analysis of permutations},
  author={Plackett, Robin L},
  journal={{Journal of the Royal Statistical Society Series C: Applied Statistics}},
  volume={24},
  number={2},
  pages={193--202},
  year={1975}
}

@book{luce1959individual,
  title={Individual Choice Behavior: A Theoretical Analysis},
  author={Luce, R Duncan},
  year={1959},
  publisher={John Wiley}
}

@article{turner2020modelling,
  title={Modelling rankings in {R}: the {PlackettLuce} package},
  author={Turner, Heather L and van Etten, Jacob and Vinoles, Concepcion},
  journal={Computational Statistics},
  volume={35},
  pages={1027--1057},
  year={2020}
}

@article{firth2005quasivariance,
  title={On the efficiency of quasi-likelihood estimation},
  author={Firth, David and De Menezes, Ren{\'e}e X},
  journal={Biometrika},
  volume={91},
  pages={65--80},
  year={2004}
}

@inproceedings{shi2025judging,
  title={Judging the judges: A systematic study of position bias in llm-as-a-judge},
  author={Shi, Lin and Ma, Chiyu and Liang, Wenhua and Diao, Xingjian and Ma, Weicheng and Vosoughi, Soroush},
  booktitle={Proceedings of the 14th International Joint Conference on Natural Language Processing and the 4th Conference of the Asia-Pacific Chapter of the Association for Computational Linguistics},
  pages={292--314},
  year={2025}
}

@inproceedings{wang2025eliminating,
  title={Eliminating position bias of language models: A mechanistic approach},
  author={Wang, Ziqi and Zhang, Hanlin and Li, Xiner and Huang, Kuan-Hao and Han, Chi and Ji, Shuiwang and Kakade, Sham and Peng, Hao and Ji, Heng},
  booktitle={Proc. ICLR},
  volume={2025},
  pages={91212--91239},
  year={2025}
}

@book{hooks1984feminist,
  title={Feminist Theory: From Margin to Center},
  author={hooks, bell},
  year={1984},
  publisher={South End Press},
  address={Boston, MA}
}

@book{fricker2007epistemic,
  title={Epistemic Injustice: Power and the Ethics of Knowing},
  author={Fricker, Miranda},
  year={2007},
  publisher={Oxford University Press}
}

@article{xu2025qwen3,
  title={Qwen3-omni technical report},
  author={Xu, Jin and Guo, Zhifang and Hu, Hangrui and Chu, Yunfei and Wang, Xiong and He, Jinzheng and Wang, Yuxuan and Shi, Xian and He, Ting and Zhu, Xinfa and others},
  journal={arXiv preprint arXiv:2509.17765},
  year={2025}
}

@article{ye2025omnivinci,
  title={OmniVinci: Enhancing Architecture and Data for Omni-Modal Understanding LLM},
  author={Ye, Hanrong and Yang, Chao-Han Huck and Goel, Arushi and Huang, Wei and Zhu, Ligeng and Su, Yuanhang and Lin, Sean and Cheng, An-Chieh and Wan, Zhen and Tian, Jinchuan and others},
  journal={arXiv preprint arXiv:2510.15870},
  year={2025}
}

@inproceedings{nainia2025beyond,
  title={Beyond BLEU: Ethical Risks of Misleading Evaluation in Domain-Specific QA with LLMs},
  author={Nainia, Ayoub and Vignes-Lebbe, R{\'e}gine and Mousannif, Hajar and Zahir, Jihad},
  booktitle={Proceedings of the First Workshop on Comparative Performance Evaluation: From Rules to Language Models},
  pages={77--86},
  year={2025}
}

@article{amini2025lfm2,
  title={Lfm2 technical report},
  author={Amini, Alexander and Banaszak, Anna and Benoit, Harold and B{\"o}{\"o}k, Arthur and Dakhran, Tarek and Duong, Song and Eng, Alfred and Fernandes, Fernando and H{\"a}rk{\"o}nen, Marc and Harrington, Anne and others},
  journal={arXiv preprint arXiv:2511.23404},
  year={2025}
}

@incollection{crenshaw2013mapping,
  title={Mapping the margins: Intersectionality, identity politics, and violence against women of color},
  author={Crenshaw, Kimberl{\'e} Williams},
  booktitle={The public nature of private violence},
  pages={93--118},
  year={2013},
  publisher={Routledge}
}

@inproceedings{bokkahalli2026seeing,
  title={From Seeing it to Experiencing it: Interactive Evaluation of Intersectional Voice Bias in Human-AI Speech Interaction},
  author={Bokkahalli Satish, Shree Harsha and Teleki, Maria and Minixhofer, Christoph and Klejch, Ondrej and Bell, Peter and Sz{\'e}kely, {\'E}va},
  booktitle={Proceedings of the Extended Abstracts of the 2026 CHI Conference on Human Factors in Computing Systems},
  pages={1--6},
  year={2026}
}

\end{document}